\documentclass[twocolumn]{aastex631}

\newcommand{\pmem}{\ensuremath{P_\mathrm{mem}}}

\shorttitle{Photometric Indicators of Cluster Relaxation}
\shortauthors{Casas et al.}

\graphicspath{{./}{figures/}{figures/gallery/}}

\begin{document}

\title {Optical Photometric Indicators of Galaxy Cluster Relaxation}

\author[0000-0002-0725-5933]{Madeline C. Casas}
\affiliation{Department of Physics, Stanford University, 382 Via Pueblo Mall, Stanford, CA 94305, USA}
\affiliation{Department of Physics, \'Ecole Normal Sup\'erieure, 24 rue Lhomond, 75005 Paris, France}

\author{Ky Putnam}
\affiliation{Department of Physics, San Diego State University, 5500 Campanile Dr, San Diego, CA 92182, USA}
\affiliation{Department of Astrophysical \& Planetary Sciences, University of Colorado, 2000 Colorado Ave, Boulder, CO 80309, USA}

\author[0000-0002-8031-1217]{Adam B. Mantz}
\affiliation{Kavli Institute for Particle Astrophysics and Cosmology, Stanford University, 452 Lomita Mall, Stanford, CA 94305, USA}

\author[0000-0003-0667-5941]{Steven W. Allen}
\affiliation{Department of Physics, Stanford University, 382 Via Pueblo Mall, Stanford, CA 94305, USA}
\affiliation{Kavli Institute for Particle Astrophysics and Cosmology, Stanford University, 452 Lomita Mall, Stanford, CA 94305, USA}
\affiliation{SLAC National Accelerator Laboratory, 2575 Sand Hill Road, Menlo Park, CA  94025, USA}

\author[0000-0003-3521-3631]{Taweewat Somboonpanyakul}
\affiliation{Kavli Institute for Particle Astrophysics and Cosmology, Stanford University, 452 Lomita Mall, Stanford, CA 94305, USA}
\affiliation{Department of Physics, Faculty of Science, Chulalongkorn University, 254 Phyathai Road, Patumwan, Bangkok 10330, Thailand}

\email{madeline.casas@ens.psl.eu,Ky.Putnam@colorado.edu, amantz@stanford.edu}


\begin{abstract}
  The most dynamically relaxed clusters of galaxies play a special role in cosmological studies as well as astrophysical studies of the intracluster medium (ICM) and active galactic nucleus feedback.
  While high spatial resolution imaging of the morphology of the ICM has long been the gold standard for establishing a cluster's dynamical state, such data are not available from current or planned surveys, and thus require separate, pointed follow-up observations.
  With optical and/or near-IR photometric imaging, and red-sequence cluster finding results from those data, expected to be ubiquitously available for clusters discovered in upcoming optical and mm-wavelength surveys, it is worth asking how effectively photometric data alone can identify relaxed cluster candidates, before investing in, e.g., high-resolution X-ray observations.
  Here we assess the ability of several simple photometric measurements, based on the redMaPPer cluster finder run on Sloan Digital Sky Survey data, to reproduce X-ray classifications of dynamical state for an X-ray selected sample of massive clusters.
  We find that two simple metrics contrasting the Bright Central Galaxy (BCG) to other cluster members can identify a complete sample of relaxed clusters with a purity of $\sim40$ per cent in our data set.
  Including minimal ICM information in the form of a center position increases the purity to $\sim60$ per cent.
  However, all three metrics depend critically on correctly identifying the BCG, which is presently a challenge for optical red-sequence cluster finders.
\end{abstract}


\section{Introduction}

\begin{table*}
  \begin{center}
  \caption{Cluster Sample}
  \vspace{-0.5ex}
  \begin{tabular}{lrrccccccccc}
    \hline
    Cluster & RA ($^\circ$) & Dec. ($^\circ$) & Redshift & Rel.\tablenotemark{a} & BCG\tablenotemark{b} & $\Delta m_{14}$ & $d_{12}$ (kpc) & CCO (kpc) & $\log_{10}|\beta|$ & $\mathrm{sign}~\beta$ & $d_{1\mathrm{X}}$ \\
    \hline\vspace{-3ex}\\
Abell   1650          &  194.6728  &  $-$1.7623   &  0.084  &  N  &  A  &  1.83  &  1.103  &  0.030  &  $-$2.69  &  +    &  0.005  \\
Abell   2142          &  239.5855  &  27.2303     &  0.089  &  N  &  R  &  1.08  &  0.179  &  0.024  &  $-$1.94  &  +    &  0.023  \\
CL      J0900+3920    &  135.0016  &  39.3479     &  0.095  &  N  &  R  &  0.82  &  0.790  &  0.266  &  $-$0.72  &  +    &  0.149  \\
Abell   2244          &  255.6773  &  34.0609     &  0.097  &  N  &  A  &  2.05  &  0.071  &  0.080  &  $-$1.43  &  +    &  0.006  \\
Abell   2034          &  227.5532  &  33.5120     &  0.113  &  N  &  R  &  1.13  &  0.053  &  0.187  &  $-$1.37  &  +    &  0.197  \\
Abell   2069          &  231.0326  &  29.8845     &  0.114  &  N  &  R  &  1.55  &  0.060  &  0.065  &  $-$1.50  &  +    &  0.036  \\
    \hline
  \end{tabular}
  \label{tab:sample}
  \end{center}
  \vspace{-1ex}
  \tablenotetext{a}{Relaxed and unrelaxed clusters according to the SPA X-ray metrics are respectively denoted Y and N \citep{Mantz1502.06020}.}
  \vspace{-1ex}
  \tablenotetext{b}{Indicates whether the adopted BCG was the most likely redMaPPer BCG candidate (R), a different redMaPPer cluster member (O) or an alternative galaxy not in the redMaPPer member list (A).}
  \tablecomments{Table~\ref{tab:sample} is published in its entirety in the machine-readable format by the Astrophysical Journal. A portion is shown here for guidance regarding its form and content.}
\end{table*}

Observations of the most dynamically relaxed clusters of galaxies enable a range of cosmological and astrophysical studies.
For these morphologically simple systems, three-dimensional profiles of gas density, temperature, pressure and entropy can be derived from X-ray data with minimal uncertainty due to projection and substructure.
In addition, through the assumption of hydrostatic equilibrium, their total mass profiles, including dark matter, can be precisely constrained.
These features enable clean investigations of cluster dark matter structure, scaling relations, ICM thermodynamics, and active galactic nucleus (AGN) feedback, compared with more dynamically and morphologically complex systems (e.g.\ \citealt{Vikhlinin0412306, Vikhlinin0507092, Arnaud0709.1561, Schmidt0610038, Mantz1607.04686, Mantz1509.01322, Darragh-Ford2302.10931}).
The precise gas and total mass constraints available for relaxed clusters also enable robust constraints on the matter density and expansion history of the Universe (e.g.\ \citealt{Allen0205007, Allen0405340, Allen0706.0033, Allen1103.4829, Ettori0211335, Ettori0904.2740, Schmidt0405374, Rapetti0409574, Mantz1402.6212, Mantz2111.09343}).

Since the launch of {\it Chandra}, the gold standard for assessing the dynamical state of massive clusters has been high spatial resolution ($<1''$) X-ray imaging of the ICM morphology.
Such observations are advantageous because the nonlinear response of the X-ray emissivity makes the presence of cool cores (which are associated with relaxation) and asymmetries due to merging or sloshing on larger scales (which are indicative of recent merger activity) clearly visible in contrast, while high resolution makes the removal of contaminating emission from point-like AGN straightforward.

While various definitions of cluster relaxation are in use, in this work we are concerned with the one employed in the most recent cosmological studies \citep{Mantz1502.06020}.
In addition to being a minority of the cluster population overall, the current sample of massive, relaxed clusters so defined is sparse at redshifts $z\gtrsim0.5$ \citep{Mantz2111.09343}.
While it is possible that the relaxed fraction evolves, numerous studies indicate that the fraction of relaxed or cool-core clusters (by various definitions) is consistent with remaining constant with redshift (e.g.\ \citealt{Santos0802.1445, Santos1008.0754,  Mantz1502.06020, McDonald1702.05094, Rossetti1702.06961, Ghirardini2106.15086}).
More likely, the sparsity of clusters in the relaxed cosmology sample at high redshifts is primarily a selection effect; the most massive clusters known at $z<0.5$ were discovered in X-ray surveys, which have an advantage in finding cool-core and relaxed systems compared with the Sunyaev-Zel'dovich (SZ) effect surveys that have discovered most of the known massive clusters at higher redshifts \citep{Mantz1502.06020, Andrade-Santos1703.08690, Rossetti1702.06961}.
Given incomplete follow-up observations of these clusters, and the fact that the cluster mass function peaks at $z\sim0.6$, the existence of even a handful of known relaxed systems at $0.5<z<1.2$ suggests that many more have yet to be identified.
Uncovering these clusters, and obtaining follow-up X-ray data, would benefit studies of the evolution of cluster astrophysics and could provide significantly improved cosmological constraints (see e.g.\ forecasts by \citealt{Rapetti0710.0440, Mantz1402.6212}).

However, obtaining {\it Chandra} observations to search for relaxed clusters, or even somewhat lower resolution XMM-{\it Newton} X-ray observations or comparable SZ imaging of the ICM with ALMA interferometry or single-dish facilities, is prohibitively expensive for the number of SZ and optical cluster detections already available at intermediate and high redshifts, let alone future surveys.
Conversely, the relatively shallow and low-resolution X-ray and SZ survey data available over the whole sky are insufficient for detailed morphological measurements on their own (though potentially extremely valuable in combination with multiwavelength data).
In this context, it is worth considering whether high-quality optical photometric data, which will soon be available for many new cluster discoveries, can be used to provide a list of relaxed cluster \emph{candidates}, for which the fraction of relaxed clusters significantly exceeds that of the general population ($\sim10$ per cent; \citealt{Mantz1502.06020}), for targeted follow-up by {\it Chandra} or other facilities.
We can generally consider such data to include the results of optical red-sequence cluster finders, as these are now routinely produced for new cluster detections regardless of the survey wavelength, at least at redshifts where appropriate  coverage is available (e.g.\ \citealt{Hilton1709.05600, Bleem1910.04121}).
In particular, these algorithms provide a list of likely cluster members, including multi-band photometry, as well as identifications of one or more candidate Bright Central Galaxies (BCGs).\footnote{As both brightness and cluster membership have associated uncertainties, we emphasize that the intended meaning of BCG in this work is \emph{bright central} galaxy and not \emph{brightest cluster} galaxy.}

In this work, we evaluate a set of particularly simple optical photometric measurements available from such data at the catalog level, potentially supplemented by basic ICM information in the form of a cluster center estimate, as may be available from X-ray or SZ survey data for particularly high confidence detections, or from brief, high-resolution snapshot observations.
In particular, we do not consider measurements requiring analysis of imaging data beyond that intrinsic to the galaxy and cluster catalogs (e.g.\ \citealt{Wen1307.0568}).

Section~\ref{sec:data} describes the data set employed here in more detail, while Section~\ref{sec:metrics} introduces the specific metrics under consideration.
We compare the performance of these metrics with X-ray morphology measurements and discuss the results in Section~\ref{sec:results} and conclude in Section~\ref{sec:conclusions}.
When converting angular to metric distances, we assume a spatially flat cosmology with a Hubble constant of $H_0=67.8$\,km\,s$^{-1}$\,Mpc$^{-1}$ and a mean matter density with respect to critical density of $\Omega_\mathrm{m}=0.308$ \citep{Planck1502.01589}.

\section{Data} \label{sec:data}
Our data set for this work consists of 100 galaxy clusters identified by redMaPPer from the Sloan Digital Sky Survey (SDSS; v6.3.1 run on the DR8 Catalog; \citealt{Rykoff1303.3562}) for which \citet{Mantz1502.06020} provide symmetry, peakiness and alignment (SPA) metrics based on {\it Chandra} X-ray imaging (Table~\ref{tab:sample}).
Briefly, the SPA algorithm produces estimates of the cluster center, its X-ray peakiness ($p$, a proxy for the presence of a cool core), symmetry and alignment ($s$ and $a$, proxies for the overall symmetry and shape of the ICM);
thresholds in $s$, $p$ and $a$ must simultaneously be exceeded for a cluster to be classified as relaxed (see \citealt{Mantz1502.06020} for details).
Within our sample, 20 clusters were classified as relaxed, and 80 as unrelaxed.
Note that the SDSS-redMaPPer catalog provides more clusters overlapping with \citet{Mantz1502.06020}, due to its larger footprint, than the more recent redMaPPer catalog from the Dark Energy Survey \citep{DES2002.11124}.
The following subsections detail augmentations of the redMaPPer catalog data used in our study.

\subsection{Missing and Misidentified BCGs} \label{sec:bcgcorrection}

In clusters with prominent cool cores, it is not unusual for redMaPPer to misidentify the BCG, often excluding the true BCG from the catalog of cluster members entirely.
This may be due to the presence of blue continuum emission and nebular emission lines associated with star formation or AGN in the BCG, causing it to depart from the red sequence, as well as the intrinsic difficulty in accurately measuring the total luminosity of an extended, diffuse galaxy with smaller galaxies in projection (e.g.\ \citealt{Crawford9903057, Rafferty0802.1864, Mantz1502.06020, Hollowood1808.06637, Kelly2310.13207}).
For cool-core clusters, given sufficiently deep imaging and filter coverage, the true BCG tends to be visually both clear and unambiguous by inspection (which is not necessarily the case for the general cluster population).

\begin{figure}
  \includegraphics[width=0.45\textwidth]{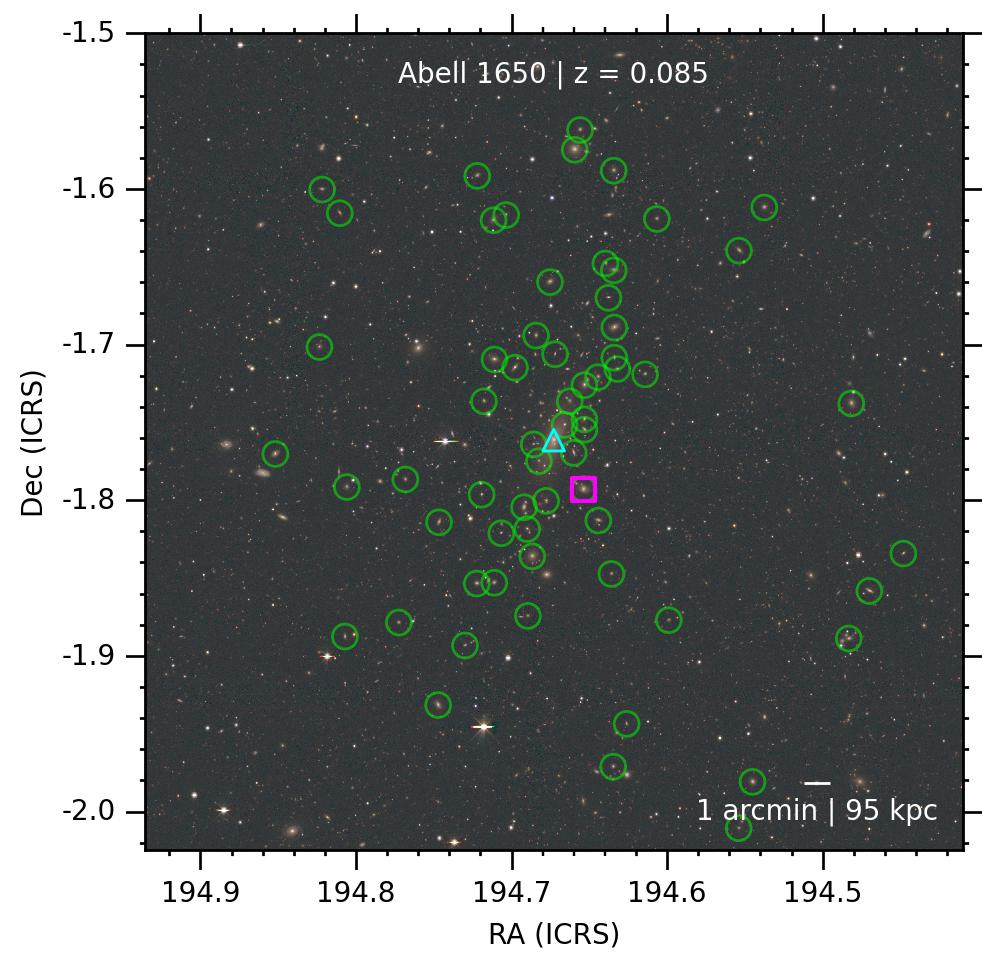}
  \caption{
    Legacy Surveys (D.~Lang, Perimeter Institute) images of clusters in our data set, ordered by increasing redshift.
    Green circles indicate galaxies assigned to each cluster by redMaPPer, with the most probably redMaPPer BCGs shown by magenta squares.
    In cases where we manually selected a different BCG, it is marked by a cyan triangle.
    Each image is centered on the cluster's X-ray position \citep{Mantz1502.06020}, and is 3\,Mpc on a side at the clusters redshift.
    The complete figure set (100 images) is available in the online Astrophysical Journal.
  }
  \label{fig:headshots}
\end{figure}

We visually inspected each of the 100 clusters in our data set, finding 21 cases where the correct BCG is absent from the list of redMaPPer cluster members entirely, and 2 where it is considered a member but is not listed as the most likely BCG candidate.
Figure~\ref{fig:headshots} shows optical images of each cluster, indicating the redMaPPer members and the cases where we manually assigned BCGs.
In one cluster, 3C288, the correct BCG lies within a mask due to the presence of a bright foreground star, and might otherwise have been identified correctly.
In all cases, BCGs chosen by inspection of the optical images coincide with the centers of the diffuse X-ray emission \citep{Mantz1502.06020}, even though the X-ray data were not taken into account at this stage.
For clusters where inspection does not reveal a clearly preferable choice, we adopt the galaxy designated by redMaPPer as the most likely BCG.

We explore the quantitative impact of missing or misidentified BCGs on various optical morphology metrics in Section~\ref{sec:cuts}.
Qualitatively, there are two immediate consequences when this occurs.
First, the galaxy designated the most likely BCG by redMaPPer may be far from the true center of the cluster, resulting in an incomplete census of cluster members and affecting estimates of the symmetry of the galaxy distribution about the BCG.
Second, metrics that involve ranking the brightness of cluster members, such as the magnitude gap, are directly affected by the absence of the brightest galaxy.

\subsection{Photometry}

\begin{figure}
    \centering
    \includegraphics[scale=0.9]{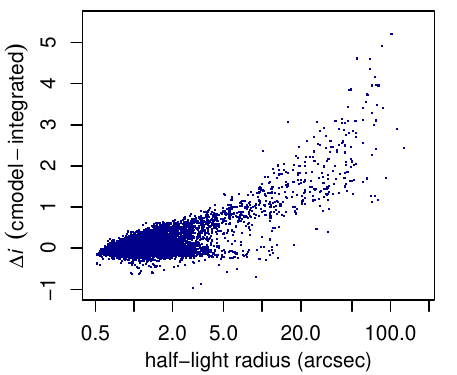}
    \caption{The difference between {\it i}-band magnitudes from the automated SDSS {\tt cmodel} measurements (used by redMaPPer) and direct integration of surface brightness profiles, as a function of size, for redMaPPer members in our sample. For larger galaxies, the direct integration systematically recovers more luminosity.}
    \label{fig:size}
\end{figure}

Another concern regarding the large, diffuse BCGs in these clusters, whose envelopes typically overlap in projection with several nearby galaxies, is the reliability of the automatic photometry provided in the public SDSS and redMaPPer catalogs, which is based on fitting various simple models to the image data.
The SDSS data release additionally provides azimuthally averaged surface brightness profiles for each galaxy, whose integral may provide a preferable aperture photometric estimate for such objects \citep{Aihara1101.1559}.
Figure~\ref{fig:size} shows that the two measurements are comparable for small galaxies ($\sim1''$ in size), while for larger galaxies the automated photometry measures less luminosity than direct integration.
Here size is defined as the radius enclosing half of the (background subtracted) light in the surface brightness profile for a source.
We adopt luminosities based on the profile integrations, and show the impact of this choice in Section~\ref{sec:cuts}.

\subsection{Cluster Membership Probability Threshold} \label{sec:Pmem}

\begin{figure}
    \centering
    \includegraphics[scale=0.9]{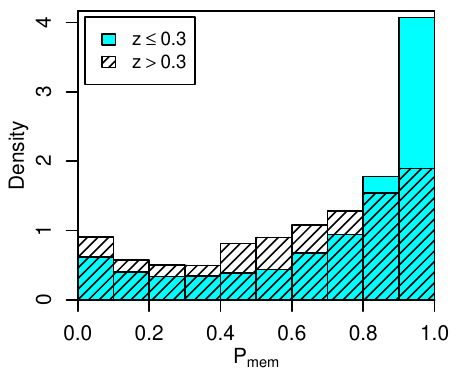}
    \caption{Histograms of redMaPPer membership probability for galaxies assigned to clusters in our sample, splitting the clusters at a redshift of 0.3. The fraction of galaxies that can be confidently assigned membership falls out to $z=0.3$, and then remains relatively constant out to $z=0.6$ (the upper limit of our sample).}
    \label{fig:pmem}
\end{figure}

RedMaPPer assigns a membership probability, \pmem{}, to each galaxy it considers a potential cluster member.
This weight depends a galaxies colors, as well as its brightness and distance from the putative cluster center \citep{Rykoff1303.3562}.
Dependent on the set of imaging bands employed and the redshift distribution of non-cluster background galaxies, the certainty with which redMaPPer can attribute individual galaxies to a cluster varies with the cluster's redshift.
Empirically, within our data set, there is a clear downward trend with cluster redshift in the fraction of galaxies with high \pmem{}, with relatively more uniform distributions of \pmem{} for clusters at $z>0.3$ (Figure~\ref{fig:pmem}).
Our baseline analysis considers only cluster members with $\pmem>0.8$, but we investigate the influence of this cut, as well as limiting the cluster redshift range, below.

\section{Optical Metrics} \label{sec:metrics}

We consider 5 morphological metrics based on photometric data (one of which requires minimal additional information about the ICM), detailed below.

\subsection{Magnitude Gap ($\Delta m_{14}$)}

Simulations have long predicted that in an isolated galaxy group or cluster, massive satellite galaxies reaching the cluster center will rapidly merge with the BCG due to dynamical friction, leading to significant growth in stellar mass and brightness of the BCG relative to other member galaxies (e.g.\ \citealt{Carnevali1981ApJ...249..449C, Mamon1987ApJ...321..622M, Barnes1989Natur.338..123B}).
This finding holds true in simulations with cosmological initial conditions, where large magnitude gaps between the BCG and other highly ranked members is correlated with earlier formation times (e.g.\ \citealt{DOnghia0505544, Dariush1002.4414, Deason1306.3990, Kundert1706.08542}).
Conversely, major mergers brings additional bright galaxies into the halo, potentially comparable to the BCG, which survive for some time before being cannibalized.
Observations furthermore associate large magnitude gaps with enhanced X-ray luminosities and larger mass concentrations at fixed total mass, both properties of relaxed halos \citep{Jones0304257, Vitorelli1708.03344}.

In this study, we define the magnitude gap, $\Delta m_{14}$, as the difference in $i$ band apparent magnitude between the first and fourth brightest cluster member.
The difference in results between first-second and first-fourth gaps (both having been used in the literature) is small, although with this sample we see slightly better performance from the latter.
Note that this metric does not explicitly depend on the assignment of the BCG, although most often the BCG is the brightest member galaxy.

\subsection{Brightest Member Distance ($d_{12}$)}

Because the timescale for dynamical friction decreases with increasing mass, the considerations above imply that the growth of the BCG comes largely at the expense of the next largest galaxies.
In other words, the second brightest member galaxy in a relaxed system is likely to be relatively distant from the BCG, so as to have avoided merging.
Cases where the second brightest galaxy appears near the BCG could simply be due to projection, or they could imply an ongoing merger in which the BCG of the smaller halo has yet to merge with that of the larger.
We therefore define our second optical metric, $d_{12}$, to be the projected distance separating the first and second brightest cluster members.
As with the magnitude gap, this definition does not explicitly rest on the designation of the BCG.
Similarly, there is a somewhat arbitrary choice of whether to compare the brightest galaxy with the second brightest or some other highly ranked galaxy; again we make this decision based on a relatively minor improvement in apparent performance within our data set.

\subsection{Central-Centroid Offset (CCO)}

Turning from tests that contrast the brightest member galaxy with others to tests of overall symmetry in the member galaxy spatial distribution, we next consider the Central-Centroid Offset.
We define this as the distance in projection from the BCG (either the highest ranked potential BCG from redMaPPer, or our manual choice) to the luminosity-weighted centroid of the non-BCG redMaPPer cluster members.
Large values of the CCO straightforwardly correspond to BCGs that are off-center compared with the other redMaPPer cluster members.

\subsection{Mirror Symmetry ($\beta$)}
We additionally employ the $\beta$ symmetry test of \citet{West1988ApJ...327....1W}, which evaluates whether there are large imbalances in mirror symmetry across the central galaxy of the cluster.
The $\beta$ statistic is defined as the an average, $\beta=\langle\beta_i\rangle$, where the index $i$ runs over non-BCG cluster members.
Here $\beta_i = \log_{10}(\tilde{d}^{~(5)}_i/d^{(5)}_i)$,
$d^{(5)}_i$ is the mean distance from galaxy $i$ to its five nearest neighbors, and $\tilde{d}^{~(5)}_i$ is the mean distance from the point opposite the $i$th galaxy (with respect to the BCG) to the five nearest galaxies.
Large absolute values of $\beta$ indicate a lack of mirror symmetry in the cluster member distribution with respect to the BCG.

\subsection{BCG to X-ray Center Distance ($d_{1\mathrm{X}}$)}

The final metric we consider combines photometric catalog data with minimal information from other wavelengths in the form of the center of the ICM.
Specifically, we use the SPA center determined from X-ray emission, measured from high-resolution {\it Chandra} data.
We define the metric, $d_{1\mathrm{X}}$, to be the projected distance between this X-ray center and the BCG.
\citet[][see also \citealt{Hudson0911.0409, Lopes1805.09631}]{Mantz1502.06020} compared this quantity with the purely X-ray SPA metrics for $>300$ clusters (a superset of our sample), finding a clear relationship between $d_{1\mathrm{X}}$ and the sharpness of the X-ray surface brightness peak.
In that work, the majority of relaxed clusters had values of $d_{1\mathrm{X}}$ that were unresolved at the resolution of the X-ray data, and essentially all clusters with $d_{1\mathrm{X}}>15$\,kpc were unrelaxed.
However, all BCG selections in their study were visually confirmed and thus potentially corrected.
In the present work, we will investigate the utility of this metric when manual BCG selection is performed only when unambiguously necessary (Section~\ref{sec:bcgcorrection}), while otherwise using the BCGs assigned by redMaPPer.

\begin{figure*}
  \centering
  \includegraphics[scale=0.99]{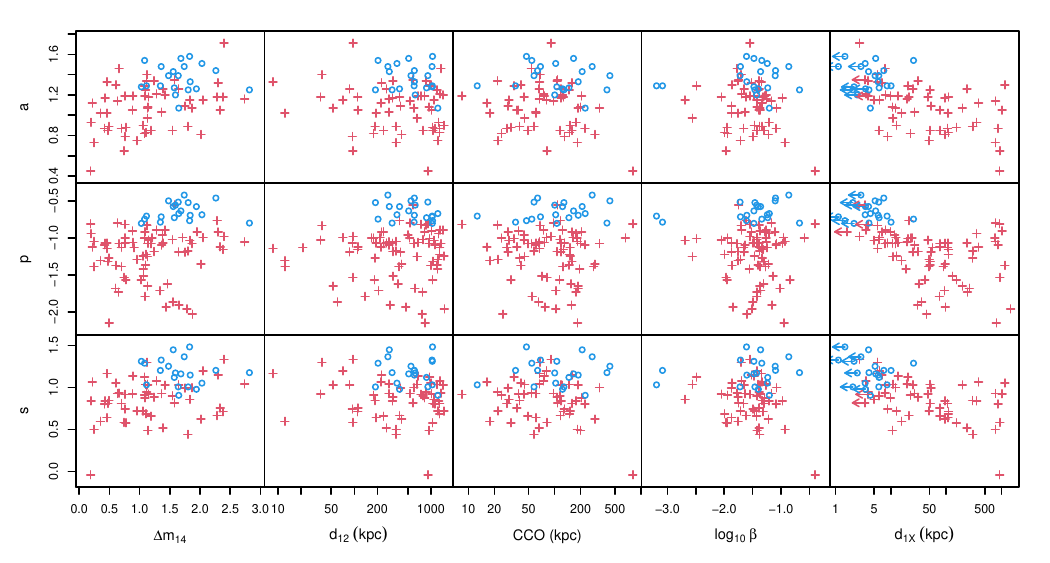}
  \vspace{-5mm}
  \caption{%
    Measured X-ray (symmetry, peakiness and alignment) and optical morphology metrics for our sample.
    Clusters classified as relaxed based on the X-ray data are shown as blue circles, and others as red crosses.
    Arrows indicate $d_{1\mathrm{X}}$ values smaller than the $1''$ resolution of the X-ray images used by \citet{Mantz1502.06020}.
  }
  \label{fig:vs_spa}
\end{figure*}

\section{Results and Discussion} \label{sec:results}

As described in the preceding sections, our baseline results incorporate manual selection of BCGs where necessary, aperture magnitudes computed from SDSS surface brightness profiles in lieu of model-fit magnitudes, and a cut of $\pmem>0.8$ for the redMaPPer cluster members.
Section~\ref{sec:vs_spa} presents these results in comparison to the SPA X-ray metrics available for our sample, while Section~\ref{sec:cuts} investigates the impact of the choices above.
In Section~\ref{sec:relaxation}, we identify simple thresholds using the optical metrics that significantly increase the fraction of relaxed clusters compared with the data set as a whole.

\subsection{Comparison to X-ray Metrics} \label{sec:vs_spa}

Figure~\ref{fig:vs_spa} compares the X-ray morphology metrics available for our cluster sample with the optical metrics defined above, distinguishing between clusters that are classified as relaxed according to the X-ray data (blue circles) and others (red crosses).
An X-ray relaxed designation requires exceeding thresholds in $s$, $p$ and $a$ simultaneously \citep{Mantz1502.06020}.
While no equivalent thresholds in the purely optical metrics appear to produce a clean sample of relaxed clusters, there are thresholds in $\Delta m_{14}$ and $d_{12}$ that might eliminate a large fraction of unrelaxed clusters.
It is less clear that the CCO and $\beta$ symmetry metrics are useful for this task, with the relaxed clusters spanning the entire observed range in both cases (albeit only due to two outliers for the latter).
Our results for $d_{1\mathrm{X}}$ resemble those of \citet{Mantz1502.06020}, with most relaxed clusters having $d_{1\mathrm{X}}<15$\,kpc (the one exception is also noted in that work), despite our having manually specified the BCG only when unambiguously necessary.

\subsection{Impacts of BCG Selection, Photometry and Membership Probability} \label{sec:cuts}

\begin{figure*}
  \centering
  \includegraphics[scale=1]{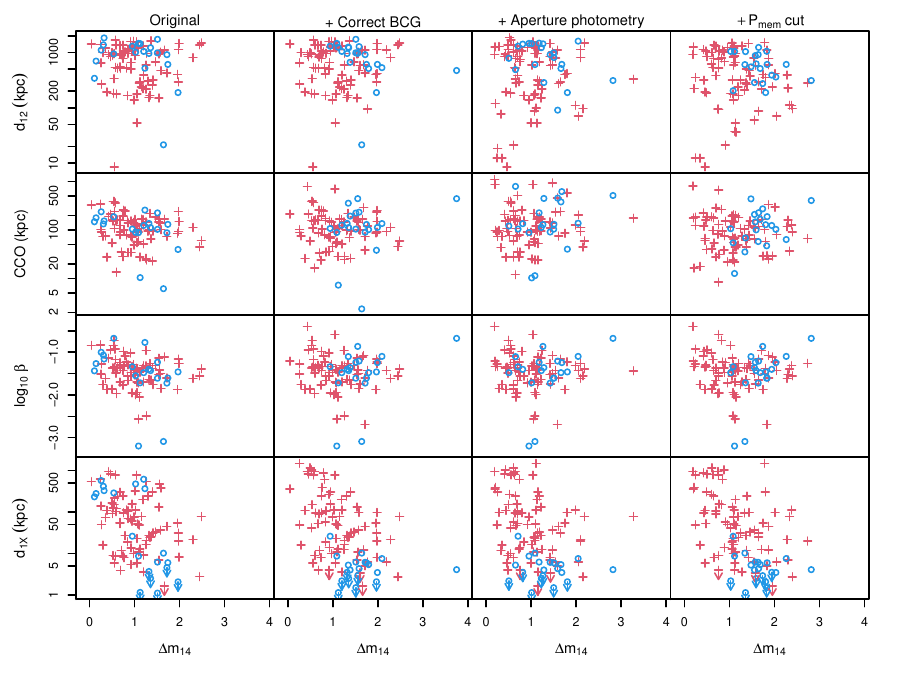}
  \vspace{-5mm}
  \caption{Optical metrics measured from the original redMaPPer catalog data, compared with various improvements (added cumulatively, left to right) discussed in Section~\ref{sec:data}. See text for details. Symbols meanings are the same as in Figure~\ref{fig:vs_spa}.}
  \label{fig:cuts}
\end{figure*}

Figure~\ref{fig:cuts} progressively shows the impacts of decisions made in Section~\ref{sec:data} on the optical morphology metrics.
The leftmost column shows $\Delta m_{14}$ against each other metric, with optical quantities computed from the original redMaPPer catalogs, specifically using their most-probable BCG selection and tabulated luminosities.
The relatively clean localization of relaxed clusters in $\Delta m_{14}$ and $d_{12}$ is not apparent in these plots; in particular, a number of relaxed clusters have physically implausible values of $\Delta m_{14}$ near zero.
Manual addition of the correct BCG to the list of cluster members, where needed, dramatically changes this picture, with all relaxed clusters having $\Delta m_{14}>1$ in the second column.\footnote{Since the added BCGs are not included in the redMaPPer tables, this step also includes a change from the redMaPPer to SDSS tabulated magnitudes. However, these are nearly identical in most cases.}
Replacing the model-inferred luminosities with integrated aperture magnitudes (third column) dramatically moves a small number of points while having a relatively small impact on the overall distribution.
Finally, enforcing a membership probability cut clearly improves the separation of relaxed and unrelaxed clusters in $\Delta m_{14}$ and $d_{12}$, but does not noticeably change the distributions of the CCO or $\beta$.
This may be because the adoption of aperture magnitudes \emph{without} removing the less likely member galaxies results in a small number of non-member foreground galaxies appearing among the most luminous considered, thus disproportionately affecting $\Delta m_{14}$ and $d_{12}$.
For $d_{1\mathrm{X}}$, the only relevant change is the manual addition of BCGs, which dramatically impacts almost half of the relaxed clusters, concentrating them at small values.

Although we do not pursue it in this work, we note that re-running redMaPPer with the cluster center fixed by fiat for the cases where we selected a BCG manually might improve the performance of the CCO and $\beta$ symmetry.
Doing so would presumably produce an improved list of cluster members in the few cases where the true BCG is relatively distant from redMaPPer's most likely BCG candidate (e.g.\ MACS~J1532, IRAS~09104).
Because they do not depend on the full list of cluster members, the metrics that we proceed with below, $\Delta m_{14}$ and $d_{12}$, would be impacted only if the original redMaPPer member list failed to include both the brightest and second brightest cluster galaxies, which is not a failure mode we have observed in this sample.

\begin{figure*}
    \centering
    \includegraphics[scale=0.9]{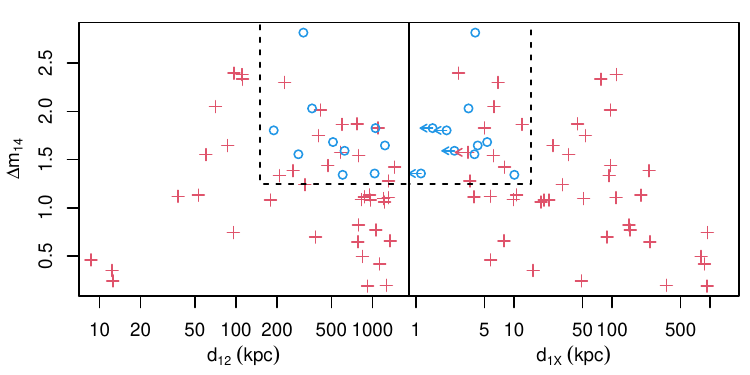}
    \caption{%
      Optical metrics as in Figures~\ref{fig:vs_spa}--\ref{fig:cuts}, restricted to clusters at $z<0.3$.
      Dashed lines denote thresholds which contain all 10 relaxed clusters in this sample.
    }
    \label{fig:finalcuts}
\end{figure*}

\subsection{Relaxed Cluster Selection} \label{sec:relaxation}

The preceding sections suggest that the most valuable purely optical metrics of those considered here are $\Delta m_{14}$ and $d_{12}$.
Indeed, Figure~\ref{fig:cuts} indicates that thresholds in these two metrics may be sufficient to identify a relatively complete sample of relaxed clusters, provided that their BCGs have been correctly identified and most non-member galaxies can be discarded.
The former requirement, at present, demands either manual inspection or analysis beyond the redMaPPer catalog construction.
The latter consideration suggests that we treat these results with caution at intermediate and high redshifts, since we saw in Section~\ref{sec:Pmem} that the certainty with which redMaPPer assigns membership to individual galaxies falls with redshift.

With this in mind, Figure~\ref{fig:finalcuts} shows the distributions of $\Delta m_{14}$ and $d_{12}$ for clusters in our sample with redshifts $z\leq0.3$.
Here, the separation of relaxed clusters is even cleaner than in Figure~\ref{fig:cuts}, with all of them satisfying the simple thresholds $\Delta m_{14}>1.25$ and $d_{12}>150$\,kpc.
Enforcing these thresholds increases the relaxed cluster fraction from $10/53\approx0.19$ to $10/23\approx0.43$. 
Note that this \emph{understates} the impact these cuts would have on an optically selected cluster sample, since the requirement of existing X-ray observation biases our data set towards containing X-ray luminous, cool core clusters. 
That is, the overall fraction of relaxed clusters is likely even lower ($\sim10$ per cent; \citealt{Mantz1502.06020}), while we do not necessarily expect the selection to significantly alter the relaxed fraction that does satisfy these cuts. 

The X-ray to BCG distance can also clearly provide helpful information, provided that X-ray data with sufficient depth and spatial resolution are available to localize the center to better than $\sim15$\,kpc.
Including a requirement that $d_{1\mathrm{X}}<15$\,kpc, in addition to the purely optical cuts above, raises the relaxed cluster fraction to $10/17\approx0.59$.

The data thus indicate that a selection of clusters using purely photometric measurements of $\Delta m_{14}$ and $d_{12}$, potentially supplemented by ICM centering information, can improve the odds of selecting relaxed clusters in this mass and redshift range by a factor of $\sim4$--$6$.
If the relaxed fraction decreases at higher redshifts and/or lower masses, the impact of these relatively simple metrics would be even greater.

\section{Conclusion} \label{sec:conclusions}

We have tested the ability of a set of simple cluster morphology metrics available from photometric survey data, specifically beginning from the redMaPPer cluster finder run on SDSS, to reproduce relaxed/unrelaxed classifications based on X-ray imaging.
We find that metrics that take advantage of the short timescale for comparably massive cluster members to merge with the BCG, namely the magnitude gap and the distance between the brightest member galaxies, are effective at isolating a complete and modestly pure ($\sim40$ per cent purity) subset of relaxed clusters in our data set.
Including minimal information about the ICM, in the form of an X-ray center measurement, increases the purity of the selection to $\sim60$ per cent.
Simple metrics based on the symmetry of the spatial distribution of photometrically identified cluster members appear to be somewhat less useful.

Some caveats apply to these results.
Primarily, we find that the effectiveness of these metrics depends on the cluster's BCG being correctly identified and included in the list of cluster members.
Relaxed clusters tend to contain exceptionally extended, bright and potentially blue (star-forming) BCGs; exactly these features make them challenging for red-sequence cluster finders to correctly associate with their host clusters.
Improving optical cluster finding techniques to identify these BCGs would have the added benefit of correctly centering such clusters for more accurate richness estimates as well as weak gravitational lensing analysis \citep{Zhang1901.07119}.
Otherwise, a second stage of analysis, either manually or automatically identifying systems where the true BCG may have been missed by a red-sequence cluster finder, is necessary.
Note that the particular fraction of problematic cases we find is dependent on the cluster sample (its selection and redshift) and the use of SDSS data rather than some other survey; however, the challenge of identifying a class of central galaxies that differs qualitatively from other cluster members is fairly generic.

Secondarily, we note that these methods are most likely to be useful at redshifts where cluster membership can be assigned with reasonable certainty, for a given survey's depth and selection of bands.
In the present work, our reliance on a relatively old archival analysis to provide ``ground truth'' in the form of X-ray relaxation determinations, limits us to studying low redshift and high mass clusters with SDSS data.
How simply our results generalize to the higher redshifts probed by the Dark Energy Survey and (in the near future) the Rubin Observatory Legacy Survey of Space and Time remains to be seen.

To maximally take advantage of the ICM-BCG offset metric requires a precision on the ICM center position of better than $\sim15$\,kpc at the cluster's redshift.
This is in principle available from either X-ray or SZ data;
although we have not tested SZ data in this work, for the clusters of interest (having relatively symmetric gas distributions centered on cool cores) the X-ray and SZ centers should in principle coincide.
Sufficient centering precision should in principle be available even from low spatial resolution X-ray (e.g.\ eROSITA or ROSAT) or SZ survey data; however, the range of cluster masses and redshifts is survey dependent.
In particular, SZ surveys with intrinsic resolution of $\sim1'$--$1.5'$ would require signal-to-noise of several tens to precisely center clusters at intermediate-to-high redshifts ($z>0.6$), limiting the utility of this metric to the most massive clusters.
With its higher spatial resolution of $\sim16''$, eROSITA X-ray data should be more broadly useful at intermediate redshifts, given $\gtrsim50$ net counts.
A second consideration when using survey data is that of contamination from AGN (X-ray or SZ) and lensed dusty, star forming galaxies (SZ) that might shift the measured center at the level of $\gtrsim1''$.
Even without such ancillary data, however, our results suggest that simple metrics from only optical imaging can provide a useful list of candidate relaxed clusters.

\begin{acknowledgments}
  We thank Eli Rykoff for helpful comments.
  MCC was supported by Stanford University's Physics Research Program.
  KP was supported by the Cal-Bridge Summer Program.
  We acknowledge support from the U.S. Department of Energy under contract numbers DE-AC02-76SF00515.
\end{acknowledgments}


Funding for SDSS-III has been provided by the Alfred P. Sloan Foundation, the Participating Institutions, the National Science Foundation, and the U.S. Department of Energy Office of Science. The SDSS-III web site is \url{http://www.sdss3.org/}.

SDSS-III is managed by the Astrophysical Research Consortium for the Participating Institutions of the SDSS-III Collaboration including the University of Arizona, the Brazilian Participation Group, Brookhaven National Laboratory, Carnegie Mellon University, University of Florida, the French Participation Group, the German Participation Group, Harvard University, the Instituto de Astrofisica de Canarias, the Michigan State/Notre Dame/JINA Participation Group, Johns Hopkins University, Lawrence Berkeley National Laboratory, Max Planck Institute for Astrophysics, Max Planck Institute for Extraterrestrial Physics, New Mexico State University, New York University, Ohio State University, Pennsylvania State University, University of Portsmouth, Princeton University, the Spanish Participation Group, University of Tokyo, University of Utah, Vanderbilt University, University of Virginia, University of Washington, and Yale University.


Figure~\ref{fig:headshots} of this work uses background images from the Legacy Surveys, whose complete acknowledgement can be found at \url{https://www.legacysurvey.org/acknowledgment/}.



\end{document}